\RequirePackage{fix-cm}
\documentclass[smallcondensed]{svjour3}     
\smartqed  
\usepackage{graphicx}
\usepackage{bm}
\begin{document}

\title{Investigation of $K^{bar}NN$ resonances with a coupled-channel Complex Scaling Method + Feshbach projection
\thanks{Details of the present study are described in our recent article \cite{ccCSM+Fesh:Dote}. }
}

\titlerunning{Investigation of $K^{bar}NN$ resonances with ccCSM + Feshbach method}        

\author{Akinobu Dot\'e \and Takashi Inoue \and Takayuki Myo}


\institute{A. Dot\'e \at
              KEK Theory Center, Institute of Particle and Nuclear Studies (IPNS), High Energy Accelerator Research Organization (KEK), 1-1 Oho, Tsukuba, Ibaraki, 305-0801, Japan \\
              \email{dote@post.kek.jp}           
           \and
           T. Inoue \at
              Nihon University, College of Bioresource Sciences, Fujisawa 252-0880, Japan
           \and
           T. Myo \at
              General Education, Faculty of Engineering, Osaka Institute of Technology, Osaka 535-8585, Japan\\
}

\date{Received: date / Accepted: date}

\maketitle

\begin{abstract}
We have investigated a prototype of kaonic nuclei, ``$K^-pp$'', with a coupled-channel Complex Scaling Method (ccCSM). Combining the ccCSM with Feshbach projection method, we can handle a coupled-channel problem effectively as a single-channel problem. By using an energy-dependent chiral SU(3)-based $\bar{K}N$ potential, the $K^-pp$ ($J^\pi=0^-$ and $T=1/2$) is obtained to be shallowly bound with the binding energy of 20-30 MeV. The mesonic decay width depends on the interaction parameters and ansatz; the decay width is ranging from 20 to 65 MeV. In case of $J^\pi=1^-$ state, no three-body $\bar{K}NN$ resonant states are found.
\end{abstract}

\section{Introduction}

In strange nuclear physics and hadron physics, kaonic nuclei (nuclear system with anti-kaons, $\bar{K}=K^-, \bar{K}^0$) have been a hot topic because they are interestingly expected to have several exotic properties, such as the formation of dense state, due to strong $\bar{K}N$ attraction \cite{AY_2002,AMDK}. To reveal such {\it expected} exotic properties involved in kaonic nuclei, we have focused on the simplest kaonic nucleus ``$K^-pp$'' which is a prototype system of kaonic nuclei. In the experimental side, new results are being reported from J-PARC experiments \cite{Kpp-ex:JPARC-E15,Kpp-ex:JPARC-E27}. Especially, J-PARC E27 collaboration has reported some signal in their $K^-pp$ search experiment using deuteron target \cite{Kpp-ex:JPARC-E27-exa2014}. In the theoretical side, the three-body system of $K^-pp$ has been investigated with various ways as summarized in Ref. \cite{SummaryKpp}.
The binding energy and decay width rather depend on approaches and employed potentials. However, all theoretical studies result that the $K^-pp$ can be bound with less than 100 MeV binding energy. Since there is $\pi\Sigma N$ threshold at 103 MeV below $\bar{K}NN$ threshold, those calculations indicate that the $K^-pp$ should be a resonant state located between the two thresholds. In addition, it is known by studies of the $\Lambda(1405)$ that the $\bar{K}N$ couples strongly to the $\pi\Sigma$  \cite{ChU:Review}. Thus, we consider that {\it 1. Resonance} and {\it 2. Coupled-channel problem} are key ingredients in the theoretical study of the $K^-pp$. We, here, employ a {coupled-channel Complex Scaling Method (ccCSM)} since this approach can simultaneously treat these two ingredients. It should be noted that the complex scaling method has greatly succeeded in studies of resonant states of stable/unstable nuclei \cite{CSM:Myo}. We have applied the ccCSM to the two-body system of $\bar{K}N$-$\pi Y$ \cite{ccCSM-NPA:Dote,ccCSM_DP:Dote}. ($Y$ means $\Lambda$ and $\Sigma$ hyperons.) Since the ccCSM is found to be a useful tool also for the study of hadronic system through those studies, we now tackle the three-body system of $K^-pp$ with the same method.

\section{Methodology}

We briefly explain our method to investigate the $K^-pp$ resonance. Here, note that the ``$K^-pp$'' called in theoretical studies means a coupled-channel system of $\bar{K}NN$-$\pi\Sigma N$-$\pi\Lambda N$ with quantum numbers of $J^\pi=0^-$ and $T=1/2$. 

Basically, we follow the usual prescription of the complex scaling method \cite{CSM:Myo} to calculate complex eigenvalues of the three-body system of $K^-pp$. The Hamiltonian for the $K^-pp$ is complex-scaled, with a complex-scaling operator $U(\theta)$ in which the coordinate and the conjugate momentum are transformed as ${\bm r} \rightarrow {\bm r}e^{i\theta}$ and ${\bm p} \rightarrow {\bm p}e^{-i\theta}$, respectively. Diagonalizing the complex-scaled Hamiltonian with a $L^2$-integrable basis functions, which are the correlated Gaussian functions \cite{CG:Suzuki} in our study, complex eigenvalues are obtained. Among those eigenvalues, the eigenvalues which are independent of the scaling angle $\theta$ correspond to the resonant states. 

\subsection{Essence of the ccCSM+Feshbach method}

The $K^-pp$ is a coupled-channel system of $\bar{K}NN$, $\pi\Sigma N$ and $\pi\Lambda N$. We treat such a multi-channel problem as a single-channel problem with help of Feshbach method \cite{Feshbach} in the complex scaling method. We call our method as a coupled-channel complex scaling method with Feshbach projection ({\it ccCSM+Feshbach method}) \cite{ccCSM+Fesh:Dote}. 

In Feshbach method, a model space ($P$ space) and outer space of the model space ($Q$ space) are assigned. Then, the Schr\"odiner equation is given as a coupled-channel equation of wave functions for $P$ and $Q$ spaces. By the elimination the $Q$-space wave function, a Schr\"odiner equation only for the $P$-space wave function $\Phi_P$ is obtained as $[ T_P + U^{eff}_P (E) ] \Phi_P = E \Phi_P$. (The operator $T_P$ is a kinetic-energy operator for the $P$ space.) Here, the effective potential for $P$ space, $U^{eff}_P (E)$, is formally written as 
$U^{eff}_P (E)=V_{PP}+V_{PQ} G_Q(E) V_{QP}$ with $G_Q(E) = (E-H_{QQ})^{-1}$, where $G_Q(E)$ is the Green function for the $Q$ space and \{$V_{XY}$\} indicates the coupled-channel potential for the $P$ and $Q$ spaces with $(X,Y)$=$P$ or $Q$. 

Certainly, a single-channel Schr\"odiner equation for the $P$ space can be obtained with the Feshbach method. However, the problem is how to represent the $G_Q(E)$ in actual calculations. We overcome this matter with a unique nature of the complex scaling method. The closure relation is proven to hold also in the CSM, including resonant states explicitly as well as bound and non-resonant continuum states. ({\it Extended Closure Relation}, ECR \cite{CSM:Myo}). The ECR is known to be well described approximately with a set of finite number of the eigenstates \{$|\chi^\theta_n\rangle$\} which are obtained by the diagonalization of a complex-scaled Hamiltonian $H^\theta$ with a Gaussian basis function; $ \sum_n \; | \chi^\theta_n \rangle \langle \tilde{\chi}^\theta_n | \; \simeq \; 1$ \cite{CSM-ECR:Myo}. 

With help of the ECR, the complex-scaled Green function $G_Q^\theta(E) = (E-H^\theta_{QQ})^{-1}$ is expressed as 
\begin{equation} 
G_Q^\theta(E) \; = \; \sum_n \frac{| \chi^\theta_{Q,n} \rangle \langle \tilde{\chi}^\theta_{Q,n} |}{E-\epsilon^\theta_n} \quad {\rm with} \quad H^\theta_{QQ} \, | \chi^\theta_{Q,n} \rangle \; = \; \epsilon^\theta_n \, | \chi^\theta_{Q,n} \rangle, \label{GQ_th}
\end{equation}  
where the energy eigenvalue $\epsilon^\theta_n$ and eigenstate $| \chi^\theta_{Q,n} \rangle$ are obtained by the diagonalization of the complex-scaled Hamiltonian $H^\theta_{QQ}$ with Gaussian basis functions. Since the non-scaled Green function for the $Q$ space is obtained from the complex-scaled Green function $G_Q^\theta(E)$ by such an inverse transformation as $G_Q(E)\;=\; U^{-1}(\theta)\,G_Q^\theta(E)\,U(\theta)$, the effective potential for the $P$ space is 
\begin{equation} 
U^{eff}_P (E) \; =\; V_{PP} \, + \, V_{PQ} \; U^{-1}(\theta)\,G_Q^\theta(E)\,U(\theta) \; V_{QP}. 
\end{equation}  
Since the $G_Q^\theta(E)$ given as Eq. (\ref{GQ_th}) is represented with Gaussian functions,  the effective potential $U^{eff}_P (E)$ is also composed of Gaussian functions. Therefore, it can be used as usual in the complex scaling method with Gaussian basis function. 

We apply this technique to the calculation of the $K^-pp$ system. Setting the $\bar{K}N$ channel as $P$ space and the $\pi\Sigma$ and $\pi\Lambda$ channels as $Q$ space, we construct an effective $\bar{K}N$ potential $U^{eff}_{\bar{K}N} (E)$. In other words, the $\pi Y$ channels are eliminated at the step of two-body calculation. The $U^{eff}_{\bar{K}N} (E)$ plugged in the three-body Hamiltonian of the $\bar{K}NN$, we solve the single-channel problem of the $\bar{K}NN$ system with the complex scaling method. 

\subsection{Treatment of the energy dependence of the effective potential \label{E-dep}}

The effective $\bar{K}N$ potential derived with the ccCSM+Feshbach method has an energy dependence. The self-consistency for $\bar{K}N$ energy has to be taken into account when bound and resonant states are considered. Such a self-consistency is dealt with a manner proposed in an earlier study \cite{Kpp:DHW}: The definition of the $\bar{K}N$ energy in the $K^-pp$ is non-trivial because the two-body $\bar{K}N$ system is a subsystem of the three-body $\bar{K}NN$ system. We define the $\bar{K}N$ energy ($E_{KN}$) in two ways by considering extreme two pictures; The anti-kaon is regarded as a field (Ansatz 1) or it is considered as a particle (Ansatz 2). Details of calculation of the $E_{KN}$ are given in Eqs. (20) and (21) in Ref. \cite{Kpp:DHW}. Also in the present study, these two ansatz are examined. We remark that this self-consistency is realized for the {\it complex} $\bar{K}N$ energy since the complex energy of the resonance pole is directly treated in the current work, while the real $\bar{K}N$ energy is considered in the earlier work with a variational approach \cite{Kpp:DHW}.

\section{Results}

\begin{figure}[t] 
\centerline{
\includegraphics[width=13pc]{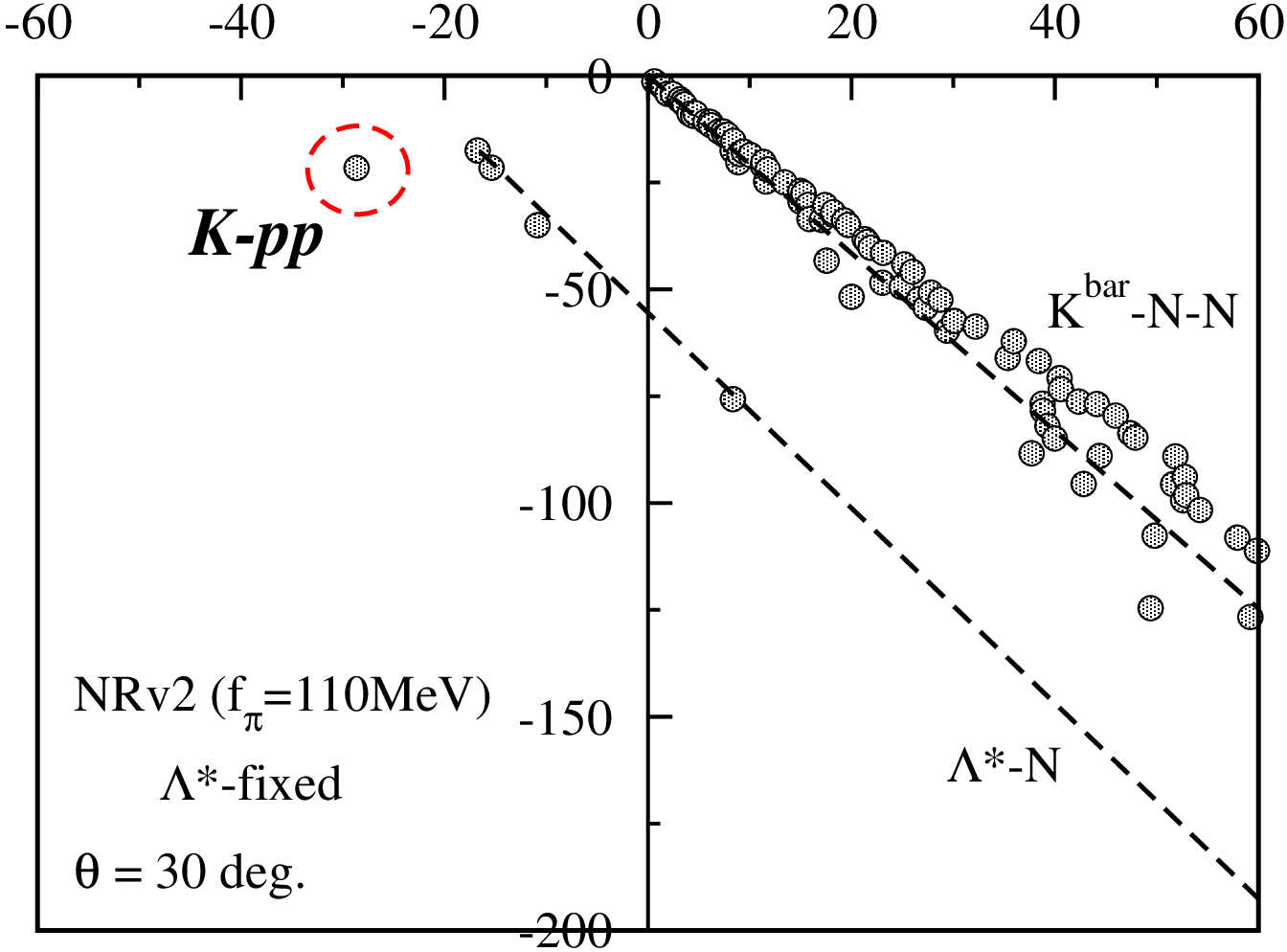}
\hspace{0.5cm}
\includegraphics[width=14pc]{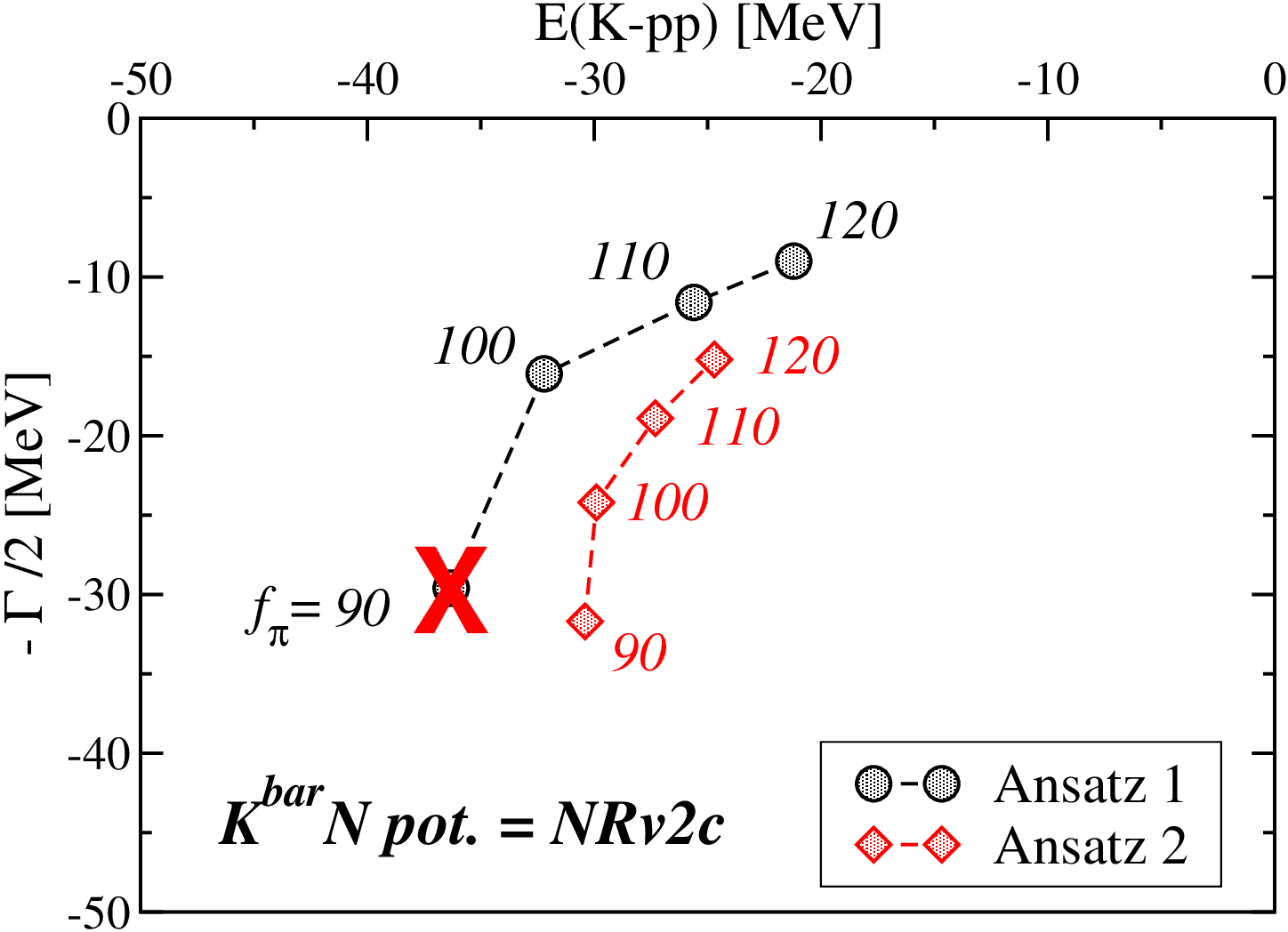}
}
\caption{\label{Fig1} (Left) Complex-energy eigenvalue distribution when $\bar{K}N$ energy for 
the effective potential is fixed to the $\Lambda^*$ energy. 
(Right) Pole position of $K^-pp$ obtained by the self-consistent calculation, 
where $f_\pi$ is varied from 90 MeV to 120 MeV and two ansatz 
for $\bar{K}N$ energy are examined.}
\end{figure} 

We show our result of the $K^-pp$ calculated with the ccCSM+Feshbach method. As a $\bar{K}N$(-$\pi Y$) potential, we use chiral SU(3)-based potentials having a single-range Gaussian form in the coordinate space, which was proposed in our previous work \cite{ccCSM-NPA:Dote}. Here, the results obtained with a version of our potentials, called NRv2c with $f_\pi=110$ MeV, are shown as a typical case. 
First, we consider the case where the $\bar{K}N$ energy in the energy-dependent $\bar{K}N$ potential is fixed to that for the $\Lambda(1405)$. In other words, the $\bar{K}N$ energy is not self-consistent in the $K^-pp$, but it is self-consistent for the isolated two-body system of $\bar{K}N$-$\pi\Sigma$ which forms the $\Lambda(1405)$ resonance. The distribution of obtained complex-energy eigenvalues is shown in the left panel of Fig. \ref{Fig1}. In the figure, the origin of the real energy axis corresponds to the $\bar{K}NN$ three-body threshold. It is known that in the complex scaling method the continuum states appear on so-called $2\theta$ line ($\tan^{-1} ({\rm Im} \, E / {\rm Re} \,E)=-2\theta$) when the scaling angle is $\theta$ \cite{CSM:Myo}. Therefore, the eigenvalues along the $2\theta$ line running from the origin indicate the scattering continuum states of $\bar{K}NN$ three-body system. There are eigenvalues on the other line. The starting point of this line is ($-16.7$, $-17.5$) MeV which is nearly equal to the complex energy of the $\Lambda^*$ \cite{ccCSM-NPA:Dote}. ($\Lambda^*$ denotes the higher pole of the $\Lambda(1405)$.) Therefore, the eigenvalues on the second line indicate the $\Lambda^* N$ two-body scattering continuum states. There is  a complex eigenvalue isolated from the two lines mentioned above, as marked with a red circle in the figure. This eigenstate means the $K^-pp$ resonance. In the case where the $\bar{K}N$ energy is fixed at that of the $\Lambda^*$, the pole energy of the $K^-pp$ resonance is found to be $(-B(K^-pp), \, -\Gamma_M/2) \, = \, (-28.6, \, -21.6)$ MeV. 

Next, we take into account the self-consistency for the $\bar{K}N$ energy in the three-body system $K^-pp$, following the two ansatz as explained in the section \ref{E-dep}. As a result of the search for self-consistent solutions, such solutions of the $K^-pp$ resonance have been successfully found. The resonance energy is obtained to be $(-B(K^-pp), \, -\Gamma_M/2) \, = \, (-25.6, \, -11.6)$ MeV for Ansatz 1 and ($-27.3$, $-18.9$) MeV for Ansatz 2. The binding energy $B(K^-pp)$ is not so dependent on the ansatz, whereas the decay width $\Gamma_M$ strongly depends on it. Compared with the earlier work of the variational calculation with a chiral SU(3)-based potential \cite{Kpp:DHW} that is a similar type to the potential used here, the present calculation gives slightly deeper binding and narrower width. 

We examine the $f_\pi$ dependence of the solutions, since the $f_\pi$ value is a parameter of our $\bar{K}N$ potential. The resonance pole of the $K^-pp$ system is found for each $f_\pi$ value when it is varied from 90 MeV to 120 MeV. (See the right panel of Fig. \ref{Fig1}.) However, since we find that the pole for $f_\pi=90$ MeV in Ansatz 1 is unstable for the scaling angle $\theta$, we discard this solution. Consequently, the binding energy is obtained to be small, 20-30 MeV. The decay width is obtained to spread widely, 20-65 MeV. We comment on the present result. It is found that both ansatz give similar binding energy. However, they give rather different decay width; Ansatz 1 tends to give extremely small decay width, compared to Ansatz 2. The Ansatz 2 seems to provide the decay width similar to that of the earlier study \cite{Kpp:DHW}.  

\begin{figure}[t]
\includegraphics[width=15pc]{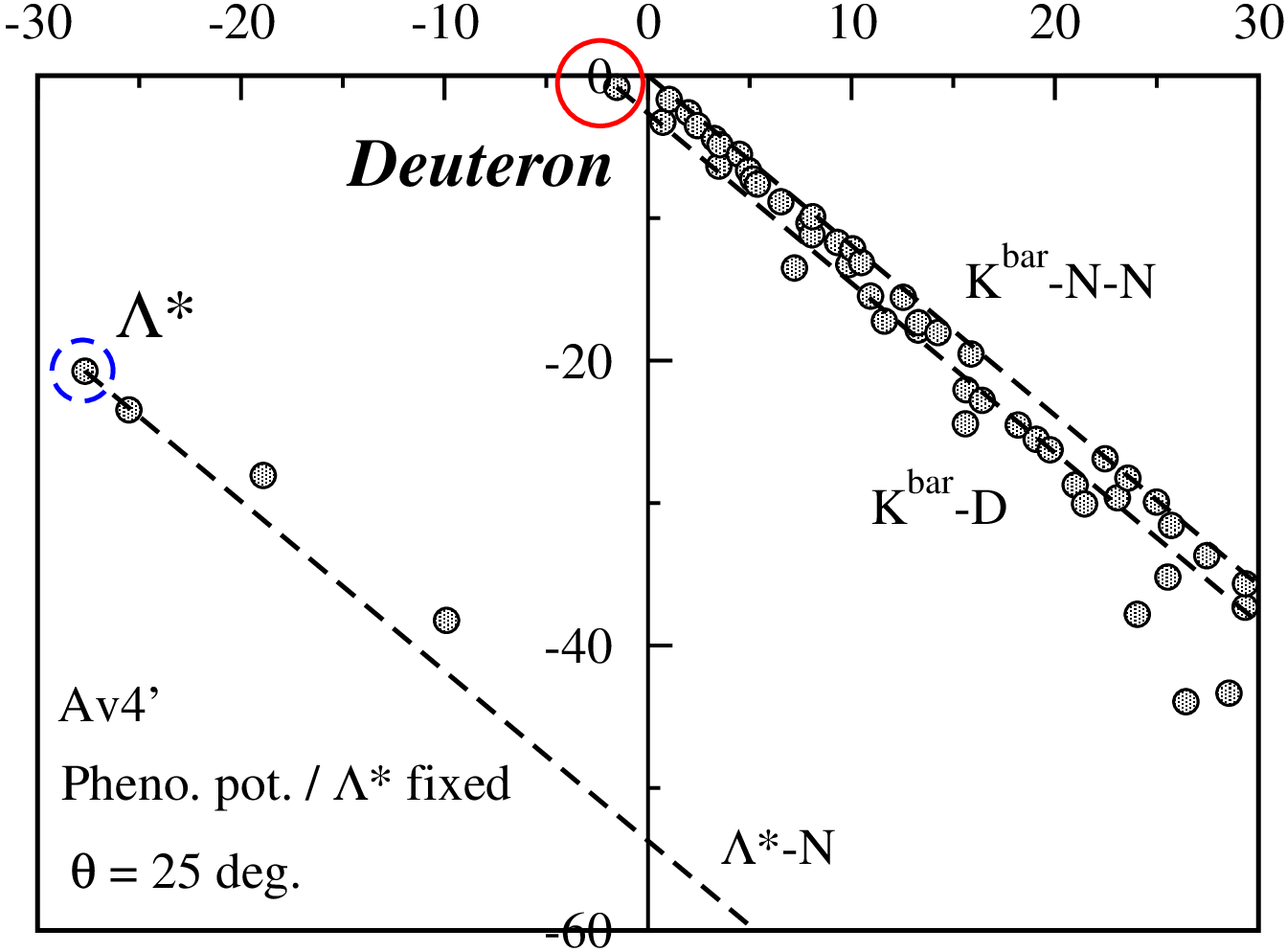}%
\caption{\label{Fig_S=1} Complex eigenvalue distribution of $\bar{K}NN$ with $J^\pi=1^-$ and $T=1/2$. Av4' $NN$ and a phenomenological $\bar{K}N$ potentials are employed. ``K$^{\rm bar}$-D'' means the scattering continuum states of $\bar{K}$ and deuteron.}
\end{figure}

As mentioned in Introduction, recently J-PARC E27 collaboration has reported their experimental result of $K^-pp$ search via $d(\pi^+, K^+)$ reaction \cite{Kpp-ex:JPARC-E27-exa2014}. We consider that this reaction excites $J^\pi=1^-$ state, not $J^\pi=0^-$ state of the $K^-pp$ that we have calculated so far, since the target of deuteron is spin 1 state and the $(\pi^+, K^+)$ reaction is expected not to flip the spin so much \cite{Discuss-Harada}. Hence, we have investigated a $\bar{K}NN$ system with the quantum numbers of $J^\pi=1^-$ and $T=1/2$ in the same way as mentioned above. Here, we set the spin and isospin of two nucleons to be $S_{NN}=1$ and $T_{NN}=0$ in the trial wave function of $\bar{K}NN$, respectively. We employ the Av4' $NN$ potential in which the tensor force is incorporated into the central force, and an energy-independent phenomenological $\bar{K}N$ potential \cite{AY_2002}. The $\bar{K}N$ energy in the effective $\bar{K}N$ potential is fixed  to be that of $\Lambda^*$ resonance for the simplicity. Fig. \ref{Fig_S=1} shows the complex eigenvalue distribution obtained with the ccCSM+Feshbach method. As marked in the figure, we can see the $\Lambda^*$ resonance, the deuteron bound state, and $\Lambda^*$+$N$ and deuteron+$\bar{K}$ scattering continuum states. There are no $\bar{K}NN$ three-body resonance states. As a result of our calculation, the $\bar{K}NN$ could not form any resonances with the quantum numbers $(J^\pi, T)=(1^-, 1/2)$ which are considered to be excited via the $d(\pi^+, K^+)$ reaction.

\section{Summary and Future Plans}

We have investigated a prototype of kaonic nuclei, $K^-pp$, with a coupled-channel Complex Scaling Method combined with Feshbach method. In the method, a coupled-channel problem is reduced to a single-channel problem to be handled more easily, by utilizing the extended closure relation in the complex scaling method. As a result of the search for the $K^-pp$ resonance pole on the complex energy plane, the binding energy and the mesonic decay width of $K^-pp$ are obtained to be 20-30 MeV and 20-65 MeV, respectively, where we use an energy-dependent chiral SU(3)-based $\bar{K}N$ potential. We consider that the $K^-pp$ is shallowly bound as suggested by earlier studies with variational methods using the same kinds of $\bar{K}N$ potentials \cite{Kpp:DHW,Kpp:BGL}. 

According to the recent report of J-PARC E27 collaboration, a bump structure is found at $\sim2260$ MeV in the $\Lambda p$ invariant-mass spectrum with two protons tagged \cite{Kpp-ex:JPARC-E27-exa2014}. This means that the binding energy is about $110$ MeV similarly to the results of the past two experiments \cite{Kpp:exp_FINUDA,Kpp:exp_DISTO}, {\it if the observed state is a $K^-pp$ state}. If this is the case, there are large discrepancy between experimental observations and theoretical results. In addition, there has been still an essential question --- ``What is the object observed in the experiments?''. To resolve these questions, further studies from both experimental and theoretical sides are necessary.
In our future plan, we will carry out the $\bar{K}NN$-$\pi\Sigma N$-$\pi\Lambda N$ coupled-channel calculation without any channel elimination. By the explicit treatment of $\pi YN$ channels we will obtain more accurate result on the $K^-pp$, and expect to clarify the role of the $\pi YN$ three-body dynamics as well as that of the $\bar{K}NN$ dynamics.

\begin{acknowledgements}
This work was developed through several activities (meetings  and workshops) at the J-PARC Branch of the KEK Theory Center. Funding: This work is funded by JSPS KAKENHI Grant Number 25400286 and partially by Grant Number 24105008. Conflict of Interest: The authors declare that they have no conflict of interest.
\end{acknowledgements}

\end{document}